\documentclass{sigchi}

\toappear{\copyright~Scott A. Hale 2014. This is the author's version of the work. It is posted here for your personal use. Not for redistribution. The definitive version was published in CHI 2015, http://dx.doi.org/10.1145/2702123.2702346.}

\usepackage{balance}  
\usepackage{graphics} 
\usepackage{times}    
\usepackage{url}      

\urldef{\wikisurvey2011}\url{https://meta.wikimedia.org/w/index.php?title=Editor_Survey_2011/Location_%26_Language&oldid=8409990} 

\makeatletter
\def\url@leostyle{%
  \@ifundefined{selectfont}{\def\UrlFont{\sf}}{\def\UrlFont{\small\bf\ttfamily}}}
\makeatother
\urlstyle{leo}

\def\pprw{8.5in}
\def\pprh{11in}

\setlength{\paperwidth}{\pprw}
\setlength{\paperheight}{\pprh}
\setlength{\pdfpagewidth}{\pprw}
\setlength{\pdfpageheight}{\pprh}

\usepackage[pdftex]{hyperref}
\hypersetup{
pdftitle={Cross-language Wikipedia Editing of Okinawa, Japan},
pdfauthor={Scott A.~Hale},
pdfkeywords={Social Media, Information Discovery, Social Network Analysis, Information Diffusion, Cross-language, Wikipedia, Multilingual},
bookmarksnumbered,
pdfstartview={FitH},
colorlinks,
citecolor=black,
filecolor=black,
linkcolor=black,
urlcolor=black,
breaklinks=true,
}

\usepackage{hypothesis}
\usepackage{booktabs}
\usepackage{graphicx}   
\usepackage[table]{xcolor}

\newcommand{\ie}{i.e.~\ignorespaces}
\hyphenation{Wiki-pe-dia}
\hyphenation{off-line}
\newcommand{\thing}{paper}
\usepackage{CJKutf8}
\usepackage{dcolumn}

\hyphenation{Wiki-Trust}
\hyphenation{Page-Rank}
\hyphenation{Wiki-Data}

\clubpenalty=10000
\widowpenalty = 10000
\interfootnotelinepenalty=10000

\title{Cross-language Wikipedia Editing of Okinawa, Japan}
\numberofauthors{1}
\author{
\alignauthor
Scott A.~Hale\\
       \affaddr{Oxford Internet Institute, University of Oxford}\\
       \affaddr{1 St Giles, Oxford, UK OX1 3JS}\\
       \email{scott.hale@oii.ox.ac.uk}
}
\date{22 September 2014}

\begin{document}

\maketitle

\begin{abstract}  
This article analyzes users who edit Wikipedia articles about Okinawa, Japan, in English and Japanese.
It finds these users are among the most active and dedicated users in their primary languages, where they make many large, high-quality edits.
However, when these users edit in their non-primary languages, they tend to make edits of a different type that are overall smaller in size
and more often restricted to the narrow set of articles that exist in both languages.
Design changes to motivate wider contributions from users in their non-primary languages and to encourage multilingual users to transfer more information across language divides are presented.
\end{abstract}




\section{Introduction}

Allowing users to contribute content in multiple languages on user-generated content platforms results in a large difference in the content available in different languages. Within Wikipedia, for example, over 74\% of concepts have an article in only one language edition, and more than 95\% of concepts appear in six or fewer languages \cite{hecht2010}. This finding is not unique to Wikipedia: on Twitter there is also surprisingly little overlap between the top domain names and hashtags used in tweets of different languages \cite{hong2011}. User-generated content platforms face a trade-off on allowing the use of multiple languages. On the one hand, increased language diversity increases the potential number of contributors as monolingual individuals from multiple languages can join. On the other hand, increased language diversity may also increase the risk of fragmenting content and users across languages, particularly if multilingual users who would have used the site in a large, international language move exclusively to smaller language editions. On question and answer platform Stack Overflow, the risk of fragmentation---that is, the risk that multiple language editions of the site would result in fewer users to any one particular language edition and therefore less high-quality answers---has been cited as one of the reasons for the platform to remain monolingual.\footnote{\url{http://blog.stackoverflow.com/2009/07/non-english-question-policy/}}

Key to the trade-off between the potential increase in other-language users and the risk of fragmentation across languages are the roles  technology and users play in facilitating the flow of information between languages. Previous research has suggested multilingual users who read and contribute content in multiple languages may share novel information between languages and broaden the scope of information available to monolingual users online \cite{eleta2012,hale-chi2014,hale-dphil-wikipedia,hecht2010}. Research to date, however, has either examined differences in content between languages \cite{bilic2014,callahan2011,hecht2010} or examined user behavior \cite{eleta2012,hale-chi2014,hale-dphil-wikipedia}, but not both. Thus, while 15\% of the active editors on Wikipedia contribute to multiple language editions \cite{hale-dphil-wikipedia}, it remains unclear what content multilingual users contribute, how much content they contribute, and how valuable the content they contribute is. These questions are important in understanding the current roles multilingual users play in transferring information between languages as well as gaining insight for designing multilingual platforms in ways that maximize cross-language information transfer.
This \thing{} starts to address this gap by examining the content contributed by multilingual users in their primary and non-primary languages on one of the largest multilingual user-generated content platforms online, Wikipedia.
The findings lead to a more in-depth understanding of the role multilingual users play on multilingual user-generated content platforms and suggest platform design strategies.

\section{Background and related work}
Each language edition of Wikipedia exhibits a self-focus: that is, each edition has, in general, more information about the regions where the language of the edition is spoken and less information about the regions where the language is not spoken \cite{hecht2010}. Even when corresponding articles for a common concept exist across multiple language editions, there is a surprising diversity in the topics covered in each language edition's article on that concept \cite{bao2012}.
For example, the Spanish-language article on psychology contains a section on important contributions from Latin America that is not found in other language editions \cite{hecht2010}.

The global nature of the Internet allows for the possibility that expatriates, language-learners, and other-culture/language enthusiasts (who Zuckerman \citeyear{zuckerman2013} has termed xenophiles) can spread information between languages on user-generated content platforms. In a one-month study of edits to the top 46 language editions of Wikipedia, Hale \citeyear{hale-dphil-wikipedia} found that approximately 15\% of active Wikipedia users edited multiple language editions of the encyclopedia. These multilingual users were distributed across all language editions, but smaller-sized editions with fewer users had a higher percentage of multilingual users compared to larger-sized editions.
The percentage of multilingual users primarily editing each language edition was found to negatively correlate with the self-focus of the 15 editions studied by Hecht and Gergle \citeyear{hecht2010}: that is, editions with more multilingual users exhibited less self-focus \cite{hale-dphil-wikipedia}.

Multilingual users are well situated to act as bridges or gatekeepers and transfer content between languages; however, past work points to a more nuanced picture of the extent to which multilingual users actually fulfill this role.
Studies of multilingual users on Twitter show they are in structural positions to act as bridges across language groups \cite{eleta2012,hale-chi2014} and that approximately 11\% of active Twitter users write in multiple languages \cite{hale-chi2014}. However, a study of multilingual users on Twitter in Switzerland, Qatar, and Quebec using LDA topic modeling found that multilingual users often focused on different topics in different languages \cite{kim2014}. This suggests these multilingual users may not be bridging across languages as much as their structural positions suggest.
A similar situation may be present in Wikipedia, where 43\% of the multilingual users studied by Hale \citeyear{hale-dphil-wikipedia} edited articles about different concepts in their primary and non-primary languages.

The first research question of this \thing{}, therefore, simply asks \researchquestion{rq:what}{what articles do multilinguals edit in their non-primary languages?}. This \thing{} compares the articles multilingual users edit in their non-primary languages with the articles edited by other users to understand the scope within which multilingual users may transfer information between languages. The results show that multilingual users edit a narrower set of articles in their non-primary languages and suggest design interventions such as cross-language content recommendation systems could broaden the scope of articles users edit in their non-primary languages.

Beyond what articles users edit in a non-primary language, this \thing{} also asks \researchquestion{rq:type}{what types of edits do multilingual users make in their non-primary languages?} in order to understand the nature and the extent of the information that multilingual users transfer between languages.
A first dimension by which to compare contributions is size. Using the meta data available from Wikipedia, Hale \citeyear{hale-dphil-wikipedia} analyzed the difference in the size of articles before and after each edit in bytes, but that measure is not the most reliable as an edit that adds a large block of text but also removes a different block of text could result in a size difference very close to zero bytes. Using the content of edits, this study calculates a more nuanced measure of edit size using the number of words added, the number of words removed, and the amount of reorganization performed.
A second dimension by which to compare contributions is the types of content changes users make in their non-primary languages. The study of 2010 Haitian earthquake blogs found that the sharing of images and videos was a large motivation for crossing language boundaries among bloggers \cite{hale-msc}, which suggests  adding, removing, or otherwise modifying images might be more prevalent in users' non-primary languages.

The final research question posed in this \thing{} asks
\researchquestion{rq:value}{how valuable are the contributions by multilingual users in their non-primary languages?}.
Given the diversity in information between languages on Wikipedia \cite{hecht2010} as well as online more generally, edits by multilingual users have the potential to introduce truly novel and valuable information and sources from one language into articles in another language. Connections across languages may serve as ``bridges'' and have been compared to the concept of ``weak ties'' in social network analysis \cite{eleta2012,hale-chi2012}, where ample scholarship has shown weak ties to be of critical importance to the spread of information, with benefits ascribed to both the individual and the system\slash{}network as a whole \cite{burt2004}.

The study of multilingual users on Wikipedia also found a positive correlation between multilingualism and the number of edits users' made in their primary languages: users more active in their primary languages were more likely to edit in multiple languages \cite{hale-dphil-wikipedia}. This suggests that multilingual users overlap to some extent with the group of very active elite or power users on Wikipedia,
on which, like on many other platforms, much of the work is done by a small percentage of very active users \cite{priedhorsky2007,kittur2007altchi}.
Monolingual studies have found that these users are responsible for a disproportionately large percentage of the content in the encyclopedia \cite{kittur2007altchi} and are overwhelmingly responsible for the content that is viewed most frequently \cite{priedhorsky2007}.
It remains unclear, however, how much content these users contribute and how long it persists when they are editing in their non-primary languages.

In order to investigate both users and content effectively, this \thing{} focuses on content and multilingual user contributions in English and Japanese in a relatively contained geographic area: Okinawa, Japan. Okinawa is an archipelago of small, sub-tropical islands home to a large number of native Japanese speakers and a large number of native English speakers.%
\footnote{The author has lived in Okinawa and can speak Japanese and English.}
Geographically closer to Taipei than Tokyo, the islands were once part of a prosperous independent kingdom built on trade in the region. After formal incorporation into Japan at the end of the 1800's, the islands were  separated from Japan at the end of World War II and administered by the United States until 1972. Since that time, the US has maintained a strong presence, with half of all US personnel (military, contractors, dependents) in Japan under the US Status of Forces Agreement located in Okinawa. This accounts for just under 50,000 individuals \cite{mofa2008}
with military facilities occupying just over 18\% of the land area of the largest and most populated island \cite{okinawa2013}.

Previous qualitative studies of Wikipedia have found differences in the editing behavior of users editing different language editions of Wikipedia---such as correlations with Hofstede's cultural dimensions \cite{pfeil2006}. These studies have not examined the roles played by users who edit in multiple languages as is done in this article, but these previous studies do underscore the importance of studying users in multiple languages before making generalizations.
Nonetheless, English and Japanese are good initial languages to study given that they are among the most-used languages online, not only on Wikipedia \cite{hale-dphil-wikipedia}, but also on Twitter \cite{hale-chi2014}. Furthermore, speakers in each language play vastly different roles in interlanguage connections \cite{hale-chi2014,hale-dphil-wikipedia}. In the one-month study of edits to Wikipedia, Japanese was a major outlier with only 6\% of the primary editors of the Japanese edition editing a second edition \cite{hale-dphil-wikipedia}. In contrast, English was very central in the cross-language movements of users: when non-English users edited a second edition, that edition was most frequently English \cite{hale-dphil-wikipedia}.

\section{Data}

Finding the subset of all articles related to a particular geographic region on Wikipedia involves a trade-off between direct relevance and completeness (or, using the terminology of information retrieval, between precision and recall). Using the Wikimedia Labs\footnote{\url{https://www.mediawiki.org/wiki/Wikimedia_Labs}} infrastructure, three article samples were extracted in October 2013. The \emph{geotag sample} included all articles with geographic information (geotags) physically placing the articles in Okinawa.%
\footnote{The bounding box used included articles between 25.75 and 26.9~E and 126 and 129~N.}
The \emph{category sample} included all articles in any category that contained the word ``Okinawa'' for the English edition or ``\CJK{UTF8}{min}{沖縄}'' (Okinawa) for the Japanese edition. Finally, the \emph{article link sample} included all articles containing a link to an article starting with ``Okinawa'' for the English edition or to an article starting with ``\CJK{UTF8}{min}{沖縄}'' (Okinawa) for the Japanese edition. All edits to each article in each sample were then downloaded from the date the article was created until October 2013 using the Wikipedia API.\footnote{\url{https://www.mediawiki.org/wiki/API:Main_page}}

The article samples were filtered to only include articles in the main, article namespace (\ie not talk pages, user pages, etc.). The article link sample was also filtered to only include articles that mentioned Okinawa in the main body text of the article (\ie not transcluded via a template to appear in a sidebar or footer). This was done so that the articles in the sample had a more substantial connection to Okinawa than just being part of a large group of articles linked together by a common portal or category such as ``Regions and administrative divisions of Japan'' or ``USAAF Eighth Air Force in World War II.'' 

Corresponding articles in the English and Japanese language editions were found using the October 2013 database dump from WikiData.\footnote{\url{http://www.wikidata.org/}}
Launched in 2012, WikiData centralizes all interlanguage links (and, increasingly, statistics and other structured data) in one location and avoids some previous issues with out-of-date or conflicting interlanguage links when the links were separately maintained in each language edition of Wikipedia. For each non-anonymous user, the Central Authorization database was queried with the username to determine if the username was a global account connected to multiple language editions.
If it was, the database for each language edition the user edited was queried to get the total number of edits per language that the user made since creating the account. The language of each user's most edited edition is referred to as the user's primary language throughout this \thing{} while the languages of any other editions edited by the user are referred to as the user's non-primary languages.
Users belonging to the (ro)`bot' group or having the `bot' template on their userpages as well as users that had been suspended for malicious editing were removed in order to focus on the behavior of good-faith, human editors.%
\footnote{The status of user accounts was checked one year after data collection in December 2014 to allow sufficient time for malicious editors to be reported and blocked. The data as well as the code used for data collection and analysis are available at \url{http://www.scotthale.net/pubs/?chi2015}.}

\subsection{Measures of edit size and value}
Users contribute value to Wikipedia in many ways. For example, Kriplean et al.~\citeyear{kriplean2008} found 42 different types of contributions to Wikipedia through an analysis of barnstars (personalized tokens of appreciation given to fellow users). The types of contributions they identified included programming tools/bots, designing templates, performing administrative functions, teaching, and leadership. Welser et al.~\citeyear{welser2011} similarly identified multiple user roles by analyzing the distribution of users' edits across the different namespaces on Wikipedia (articles, article talk pages, user pages, etc.).

In order to have one quantitative measure of the value of edits to articles, this \thing{} uses edit persistence, following past work analyzing Wikipedia in one language \cite{adler2008-assigningtrust,adler2007,adler2008-measuring,priedhorsky2007}. In particular, this \thing{} uses the algorithms developed by Adler et al.~\cite{adler2008-assigningtrust,adler2007,adler2008-measuring} for their WikiTrust content-driven reputation system for Wikipedia to compute the extent to which each edit survives through the next six revisions to the article (edit persistence). WikiTrust accounts for some of the complexities of Wikipedia including differences created by rewording and rearranging text, and it also correctly attributes text that is deleted and then restored to the first author who contributed it (rather than to the author who restored it) \cite{adler2007}.
Edit persistence is calculated on a word-by-word basis for the next six revisions to each article so that users get partial credit for an edit even when part of their edit is subsequently changed or removed.

WikiTrust also computes a more detailed measure of edit sizes than is reported directly from Wikipedia. The edit sizes reported by Wikipedia simply measure the difference in page size (in bytes) before and after an edit. In contrast, the sizes computed by WikiTrust and used in this \thing{} are determined by the number of words added, deleted, changed, or moved. New words and deleted words contribute one point each, replacement words contribute 0.5 points each, and moving a word a fraction $x$ of the normalized page length ($0<x<1$) contributes $x$ of a point \cite{adler2007}.

WikiTrust was developed and tested on languages with spaces between words; however, Japanese is written without spaces between words. Therefore, the Japanese text was first preprocessed to add spaces between words with mecab, an open-source library for text segmentation, part-of-speech tagging, and morphological analysis of Japanese text.\footnote{\url{https://code.google.com/p/mecab/} (in Japanese)}
The WikiTrust algorithm was then run twice: once over the articles in the English article link sample and once over the articles in the Japanese article link sample. The reputation scores were computed only on the basis of users' edits in the data samples, and not in the full multi-terabyte dumps of all of Wikipedia articles.

\section{Results}

This section begins by describing the three article samples in order to understand the article landscape within which Wikipedia users were editing. It then addresses each of the three main research questions in turn:
\begin{enumerate}
\setlength{\itemsep}{1pt}
\setlength{\parskip}{1pt}  
	\item \PrintRQ{rq:what}
	\item \PrintRQ{rq:type}
	\item \PrintRQ{rq:value}
\end{enumerate}

\begin{table}
\begin{center}
\begin{tabular}{lrrrrr}
\toprule
	Sample		&	en-only	& ja-only	& Both\\
\midrule
	Geotag		&	52		&	185		&	152\\ 
	Category		&	156		&	2,819	&	707\\
	Article link	&	3,411	&	9,984	&	5,567\\ 
\bottomrule
\end{tabular}
\caption{The number of unique concepts in each sample. The majority of concepts have an article either only in the English edition or only in the Japanese edition (en-only or ja-only), while a smaller number of concepts have articles in both the English and Japanese editions (Both).}
\label{tbl:sample_stats}
\end{center}
\end{table}

All three article samples (geotag, category, and article link) show differences in the concepts related to Okinawa covered in the Japanese and English editions of Wikipedia. Consistent with previous research \cite{hecht2010}, there are more concepts with an article in only one language (either Japanese or English) than concepts with articles in both languages (Table \ref{tbl:sample_stats}). The three samples have substantial overlap---all but a small handful of articles in the geotag sample are present in the category sample, and the article link sample contains all the articles in both the geotag sample and the category sample.

In order to investigate the differences between the editions further, the inter-article links that connect articles together within each language were used to construct two networks for each sample. One network for the Japanese edition with each Japanese article in the sample as a node, and one network for the English edition with each English article as a node. Edges in both networks were the inter-article links between articles in the same language edition. Nodes were then ranked using the PageRank method \cite{page1999}. This algorithm, also used by Google, ranks nodes by the number of links to them weighted by the PageRanks of the nodes from which the links originate.

Within the geotag sample, the top-ranked English-only articles were mainly about US military facilities, equipment, and historic battles, while the top-ranked geotagged Japanese articles included a variety of parks, tourist areas, ports/terminals, and a shrine, reflecting that Okinawa is a Japanese tourist hot spot.

The top-ranked articles within the category and article link samples were very similar to each other. The top-ranked articles appearing only in English in the category and article link samples included many articles related to karate, which started in Okinawa. The top-ranked articles appearing only in Japanese in these samples included historic articles (articles on the reversion of Okinawa from the US to Japan in 1972) as well as articles about Okinawa-based media and transit companies.

Given the substantial overlaps between the samples, the remainder of this \thing{} uses only the article link sample to investigate the roles of multilingual users in the spread of content between the Japanese and English language editions. The article link sample has the benefits that an article link sample can be formed for any seed set of articles and that article links have been better studied on Wikipedia \cite{milne2008} than categories. In particular, Bao et al.~\cite{bao2012} examined the frequency with which different language editions were missing article links (that is the frequency with which a concept was mentioned without a link to the article about the concept) and found that the English and Japanese editions were missing possible article links at similar frequencies.

\begin{table*}
\begin{center}
	\begin{tabular}{llrrrrrrrrrrrrrrrr}
\toprule
	&		&	\multicolumn{2}{l}{Total users}	&	\multicolumn{2}{l}{Total articles edited}	&	\multicolumn{3}{l}{Articles edited per user}	&	\multicolumn{3}{l}{Edit size per user (log)}\\

&	& \multicolumn{1}{l}{Count}	& \multicolumn{1}{r}{\%}	&\multicolumn{1}{l}{Count}	& \multicolumn{1}{r}{\%}	&	\multicolumn{1}{l}{Median} & \multicolumn{1}{l}{Mean} & \multicolumn{1}{l}{SD}	&	\multicolumn{1}{l}{Median} & \multicolumn{1}{l}{Mean} & \multicolumn{1}{l}{SD}\\
\midrule

\multicolumn{2}{l}{\textbf{English edition}}\\
& Anonymous	&	192,839	&	73.4\%	&	216,840	&	46.15\%	\\
& Local account	&	15,008	&	5.7\%	&	58,689	&	12.49\%	&	1	&	3.91	&	21.79	&	1.79	&	1.97	&	1.93\\
& Pri. English	&	50,038	&	19.0\%	&	179,951	&	38.30\%	&	1	&	3.60	&	20.74	&	1.94	&	2.08	&	1.97\\
& Pri. Japanese	&	466	&	0.2\%	&	1,488	&	0.32\%	&	1	&	3.19	&	7.32	&	1.16	&	1.38	&	1.67\\
& Pri. Other	&	4,341	&	1.7\%	&	12,911	&	2.75\%	&	1	&	2.97	&	16.44	&	0.47	&	1.13	&	1.71\\
& Totals	&	262,692	&	100.0\%	&	469,879	&	100.0\%	&	1	&	3.62	&	20.67	&	1.84	&	2.00	&	1.96\\

\multicolumn{2}{l}{\textbf{Japanese edition}}\\
& Anonymous	&	372,852	&	88.4\%	&	717,608	&	62.74\%	\\
& Local account	&	9,945	&	2.4\%	&	109,765	&	9.60\%	&	2	&	11.04	&	47.84	&	3.09	&	2.95	&	1.81\\
& Pri. English	&	558	&	0.1\%	&	5,531	&	0.48\%	&	1	&	9.91	&	58.47	&	0.96	&	1.55	&	1.95\\
& Pri. Japanese	&	37,191	&	8.8\%	&	301,980	&	26.40\%	&	1	&	8.12	&	43.97	&	3.00	&	2.91	&	1.83\\
& Pri. Other	&	1,174	&	0.3\%	&	8,954	&	0.78\%	&	1	&	7.63	&	57.44	&	0.18	&	1.07	&	1.76\\
& Totals	&	421,720	&	100.0\%	&	1,143,838	&	100.0\%	&	1	&	8.72	&	45.35	&	2.97	&	2.87	&	1.85\\

\bottomrule
\end{tabular}
		\caption{User counts, articles edited, and edit sizes. The primary (pri.) language of a user with a global account is the language of the most-edited edition of Wikipedia.}
		\label{tbl:editCounts}

\end{center}
\end{table*}

\begin{figure}
\begin{center}
	\includegraphics[width=\columnwidth]{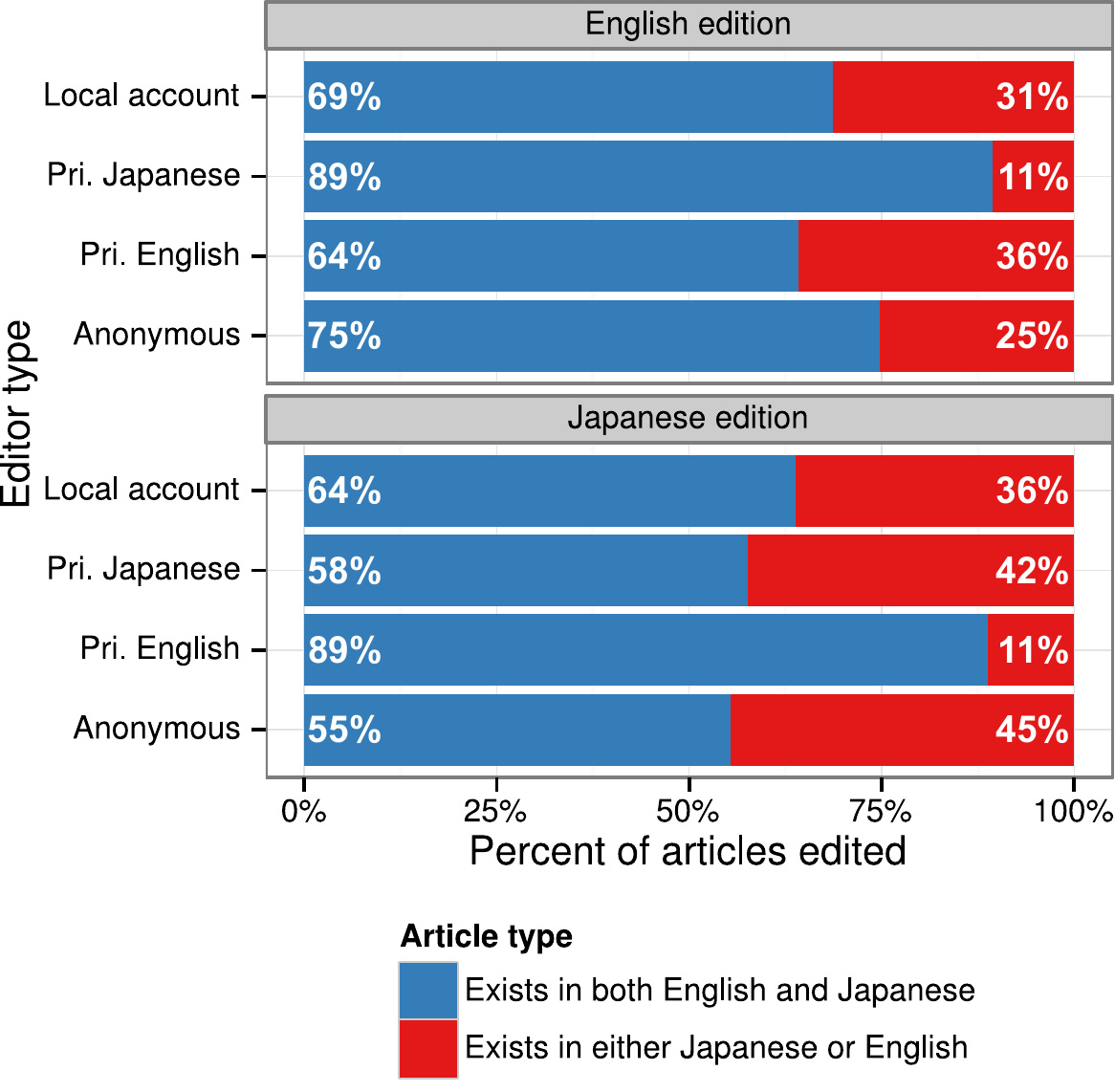}
	
	\caption{English users editing the Japanese edition are far less likely than other users to edit articles that only appear in Japanese. Similarly, Japanese users editing the English edition are far less likely than other users to edit articles that only appear in English.}
	\label{fig:article-types}
\end{center}
\end{figure}

\subsection{Article selection}
This subsection examines what articles users edited, with a particular emphasis on what articles multilingual users edited in their non-primary languages in order to answer the first research question. Most editors were anonymous, had local accounts, or had global accounts that primarily edited the language edition in question (Table \ref{tbl:editCounts}). The prevalence of edits by anonymous users is consistent with previous research analyzing Wikipedia and not a peculiarity of this sample \cite{anthony2009}. It is difficult to calculate per-user statistics for anonymous users since IP addresses change over time and multiple users may edit from the same IP address, but relevant statistics for anonymous users are presented where possible given the size of this group.

A small number of registered users primarily edited either the Japanese or English edition, but also edited the opposite edition: 558 primary editors of the English edition edited articles in the article link sample from the Japanese edition, and 466 primary editors of the Japanese edition also edited articles in the article link sample from the English edition.

The articles users choose to edit reflect how different groups of users distribute their time and energy across articles.
Although the previous section showed that most concepts had articles in only one language, the majority of edits by all users were to concepts with articles in both languages (Figure \ref{fig:article-types}). Even so, the edits by multilingual users writing in a non-primary language were significantly more concentrated on concepts with corresponding articles in both languages compared to the edits of other users writing in each language.%
\footnote{The difference in means between any two groups within either edition is significant with $p<0.001$.}

\begin{table*}\centering
\begin{tabular}{@{\extracolsep{5pt}}l D{.}{.}{-3} D{.}{.}{-3} D{.}{.}{-3} D{.}{.}{-3} }
\toprule
& \multicolumn{2}{c}{\# of Japanese users editing English} & \multicolumn{2}{c}{\# of English users editing Japanese} \\
& \multicolumn{1}{c}{Estimate} & \multicolumn{1}{c}{(Standard error)} & \multicolumn{1}{c}{Estimate} & \multicolumn{1}{c}{(Standard error)} \\
\midrule

Exists in both languages & 0.641^{***} & (0.024) & 3.285^{***} & (0.034) \\
Total number of editors & 0.001^{***} & (0.0001) & 0.003^{***} & (0.0001)\\
PageRank & 0.014^{***}  & (0.0005) & 0.245^{***} & (0.006) \\
Number of images & 0.003^{***}  & (0.001) & 0.054^{***} & (0.002)\\
Number of external links & 0.001^{***}   & (0.0003) & -0.0003 & (0.0004)\\
Constant & 0.008 & (0.015)  & 0.029 & (0.019) \\
\midrule
Observations & \multicolumn{2}{c}{5,441} & \multicolumn{2}{c}{14,825} \\
Adjusted R$^{2}$ & \multicolumn{2}{c}{0.348} & \multicolumn{2}{c}{0.572} \\
Residual Std. Error & \multicolumn{2}{c}{0.849 (df = 5435)} & \multicolumn{2}{c}{1.828 (df = 14819)} \\
\bottomrule
\multicolumn{5}{r}{$^{*}$p$<$0.1; $^{**}$p$<$0.05; $^{***}$p$<$0.01} \\
\normalsize
\end{tabular}
\caption{Linear regression results fitting the number of primary Japanese users editing each English article and the number of primary English users editing each Japanese article.}
\label{tbl:regression}
\end{table*}

The finding that multilingual users mostly edit articles with corresponding articles in their primary languages is confirmed by a linear regression (Table~\ref{tbl:regression}), which also shows the articles users edited in their non-primary languages tended to have more overall edits\slash{}editors, more images, and have higher PageRank scores computed as described earlier. The articles Japanese users edited in English also tended to have more links to external sources, but the number of links to external sources was not significantly associated with the number of English users editing an article in Japanese.

These results give a more nuanced understanding of the article selection behavior of multilingual users. Data on the articles that users viewed is not available, and thus it is not possible to say whether multilingual users read (but did not edit) articles in their primary languages before editing the corresponding articles in their non-primary languages.
The editing data, however, clearly shows that while multilingual users in this dataset edit similar proportions as other user groups of one- and two-language concepts in their primary languages, they disproportionately edit a smaller amount of one-language concepts in their non-primary languages.

\subsection{Types of contributions}
This subsection addresses the second research question on the size and type of edits multilingual users make. Setting aside anonymous users and local accounts to compare users who have global accounts to one another, a finding in both editions is that those users who edited articles in both Japanese and English in the dataset were very active on their primary language editions. Considering users with global accounts who primarily edited the English edition, about one percent of these users also edited the Japanese edition. However, this group of users was responsible for six percent of all edits to the English edition made by English users. Similarly, Japanese users who also edited the English edition were also about one percent of all Japanese users with global accounts, but they made 13\% of all edits by Japanese users in the Japanese edition.
\begin{figure*}
	\begin{center}
		\begin{center}
		\includegraphics[width=0.45\textwidth]{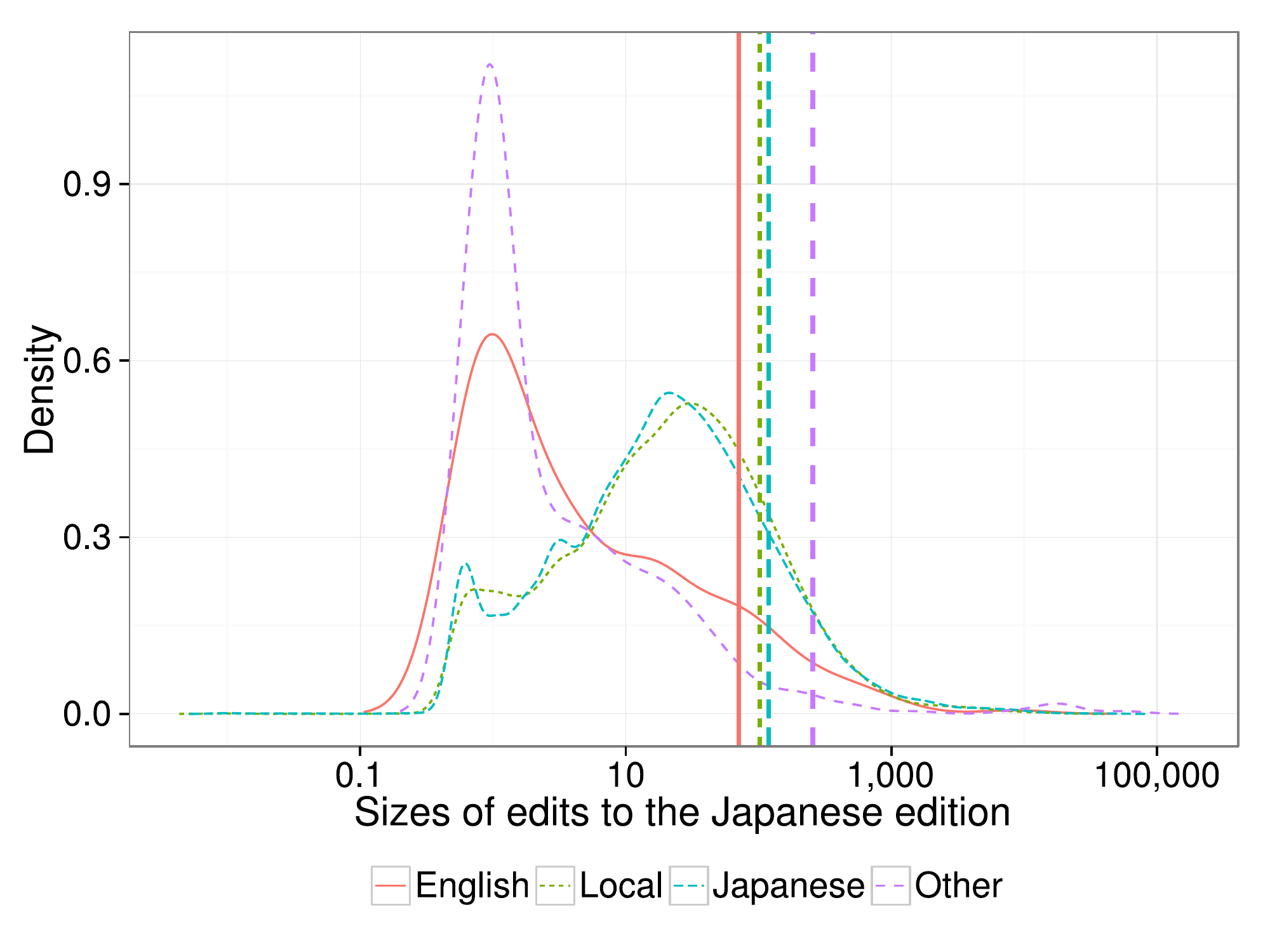}%
		\hspace{2em}%
		\includegraphics[width=0.45\textwidth]{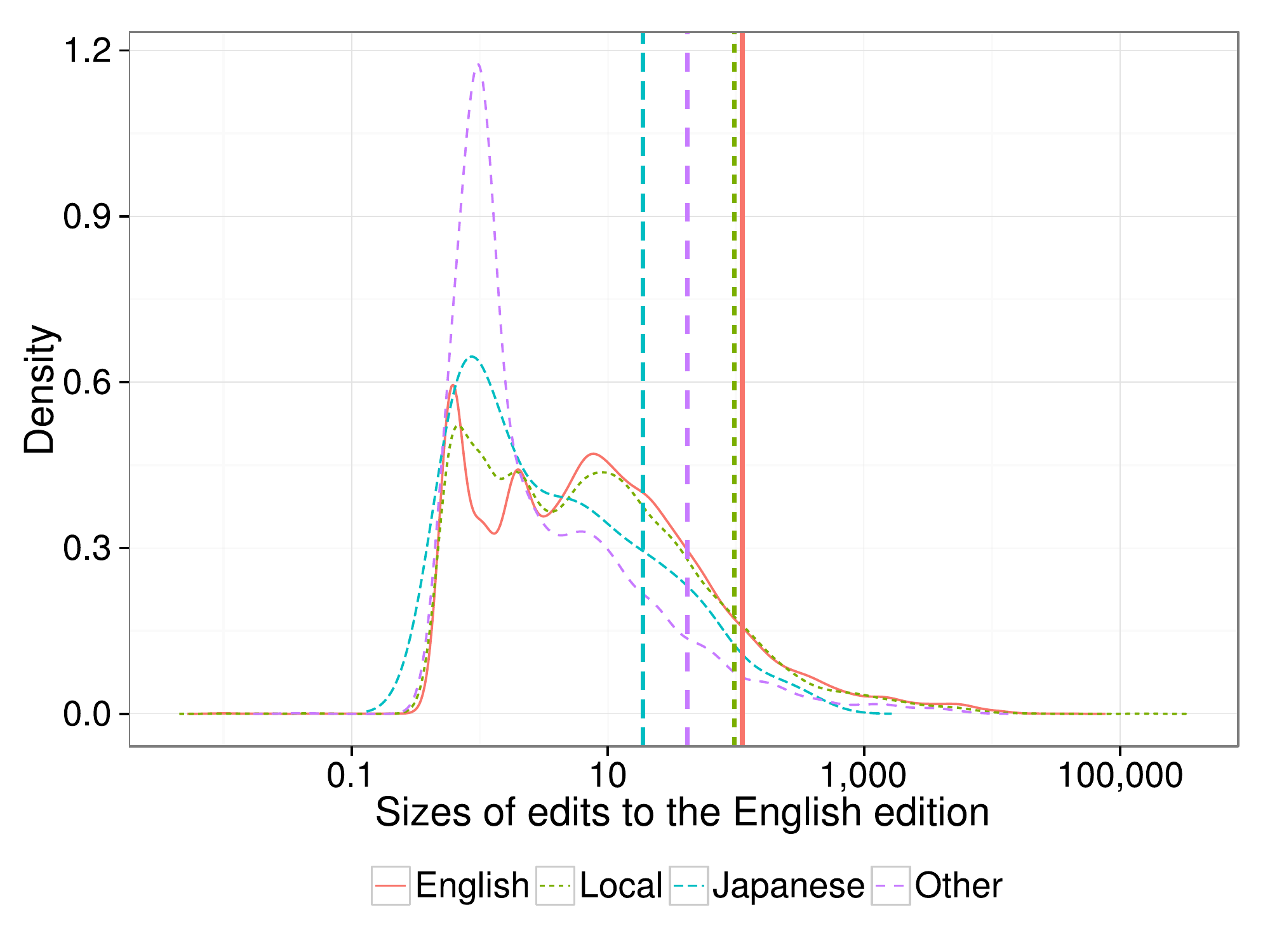}
		\caption{Density plots for non-anonymous users editing articles in the Japanese (left) and English (right) editions grouped by their primary language editions. Vertical lines indicate distribution means.}
		\label{fig:delta_group}
		\end{center}
	\end{center}
\end{figure*}

While multilingual users were very active in their primary languages, a smaller percentage of each user's total edits were to each user's non-primary languages. Overall, 14\% of all edits by multilingual users were to their non-primary languages, which is higher than the prior work that found only 2.6\% of edits across all language editions for one month were from users writing in their non-primary languages \cite{hale-dphil-wikipedia}. This could be due to the focus on a geographic region with large numbers of English and Japanese speakers and/or the longer time frame of the analysis.

\subsubsection{Edit sizes}

Users who edited both the Japanese and English editions have two sets of scores for their edit sizes and edit persistence. One set for the English edition and one set for the Japanese edition. First looking at the scores for users' primary language editions, a consistent finding in both Japanese and English is that those users who edited both editions made significantly larger edits than the users who only edited one edition.%
\footnote{Given the heavy-tailed distribution of average edit sizes, the results are reported after being transformed with the logarithm.}
Japanese users who also edited the English edition made larger sized edits in Japanese compared to Japanese users who did not edit the English edition (median 3.9 vs.~3.0, mean 3.8 vs.~2.9, sd both 1.8, $p<0.001$).
Likewise, English users who also edited the Japanese edition made larger sized edits in English compared to English users who did not edit the Japanese edition (median 2.4 vs.~1.9, mean 2.5 vs.~2.1, sd 1.9 and 2.0, $p<0.001$).

While multilingual users made larger sized edits in their primary languages, the analysis of their edits in their non-primary languages reveals a very different picture (Figure \ref{fig:delta_group} and Table \ref{tbl:editCounts}).
The sizes of edits to the Japanese edition by English users were significantly smaller than the sizes of edits to the Japanese edition by Japanese users ($p<0.001$). Similarly, the sizes of edits to the English edition by Japanese users were significantly smaller than the sizes of edits to the English edition by English users ($p<0.001$).

\subsubsection{Content changes in primary and non-primary languages}

\begin{table}\centering
	\begin{tabular}{lrrrrr}
\toprule	
Edit category	&	\multicolumn{2}{c}{Pri.\ lang.}	&	\multicolumn{2}{c}{Non-pri.\ lang.}	&	p-val\dag\\
\midrule
Addition	&	97	&	31\%	&	47	&	26\%		&	0.25\\
Maintenance	&	103	&	33\%	&	44	&	24\%	&	0.04\\
Deletion/Reversion	&	37	&	12\%	&	11	&	6\%	&	0.03\\
Image-related	&	27	&	9\%	&	32	&	18\%		&	0.01\\
Interlanguage links	&	8	&	3\%	&	32	&	18\%&	0.00\\
Change	&	65	&	21\%	&	34	&	19\%		&	0.62\\
\midrule
Total edits\ddag	&	315	&		&	181	&		&		\\
\bottomrule
	\end{tabular}
\caption{Exploratory, qualitative coding of edits in users' primary languages (pri.\ lang.) and non-primary languages (non-pri.\ lang.).\newline
\dag p-values are for two-tailed t-tests on difference of percentage means. \ddag Some edits are assigned to multiple categories and, therefore, the column sums are greater than the total number of edits reported.}
\label{tbl:qual}
\end{table}

In order to better understand the gap between the edit sizes of users in their primary and non-primary languages, a small subset of edits was explored qualitatively. A random set of 70 users who had edited both editions was chosen: 35 users who primarily edited the English edition and 35 users who primarily edited the Japanese edition. Up to five edits from each edition were randomly chosen for a total of up to 10 edits from each user. Despite the measures to remove (ro)bots described in the data section, qualitative analysis revealed that one randomly chosen user was a bot: this user was replaced with another randomly chosen user. Not all users had five edits in each edition; so, 496 edits were reviewed in total. This set included edits by English users to the English edition (145 edits) and the Japanese edition (96) as well as edits by Japanese users to the English edition (85) and the Japanese edition (170). The findings suggest that users made different types of contributions in their primary and non-primary languages, which may account for the differences in the computed size of their edits.

Edits made to articles (but not to talk pages, etc.) were examined in order to understand the contributions users made to articles in their primary and non-primary languages. After consulting previous literature \cite{kriplean2008,pfeil2006}, initial codes were created through an emergent coding of a subset of the data. These were refined into six (non-exclusive) categories, and the full sample was systematically coded using these categories. Each edit was classified as making an addition (adding new text or references to an existing article or creating a new article), as maintenance (adding, removing, or adjusting templates, categories, links in a ``See Also'' section, or whitespace changes that did not alter text), as deletion/reversion (reverting an edit or deleting text from an article), as image-related (adding, altering, or removing an image), as altering interlanguage links, and/or as change (edits that changed existing text such as correcting spelling errors or updating facts that had changed like the latest winner of an annual sports tournament).

There was a significant difference between the types of edits users made in their primary languages compared to the types they made in their non-primary languages ($\chi^2=48, p<0.001$,  Table \ref{tbl:qual}). Users made significantly more deletions/reversions and maintenance edits in their primary languages compared to in their non-primary languages. On the other hand, users made significantly more image-related edits and added/removed significantly more interlanguage links in their non-primary languages compared to in their primary languages. The findings related to interlanguage links are no longer applicable to Wikipedia as these links are now maintained separately within WikiData. Nonetheless, it is noteworthy that the task of locating a related article and linking it across languages was motivating enough for some users to edit a foreign language edition.

The proportion of addition edits and change edits did not differ significantly between users' primary and non-primary languages. 
Overall, 15\% of the edits made by Japanese users to the English edition concerned fixing incorrect romanizations of Japanese words and/or adding Japanese characters for terms. These types of language-specific edits that are easy for native speakers but harder for non-native speakers illustrate both the value of cross-language collaboration and also why these users may have been making edits of different types in their primary and non-primary languages.

\subsection{Value of edits}

As stated in the Data section, there are many ways in which users contribute value to Wikipedia, but a common, quantitative measure on which to compare the many different types of contributions is how much of each edit is retained through subsequent revisions to the article. Using the WikiTrust algorithms, the next six revisions after each edit were examined to compute how much of the edit was retained (persisted) through these revisions. Each edit was given a normalized score from -1 (edit completely removed) to 1 (edit completely retained).
Comparing the mean edit persistent scores showed that the text from edits made by non-primary editors survived at a similar rate to the text from edits made by users who primarily edited each edition.%
\footnote{The average score for Japanese users editing the English edition (median 0.44, mean 0.38, sd 0.58) was higher but not significantly different from the average score for English users editing the English edition (median 0.47, mean 0.35, sd 0.64, $p=0.46$). Similarly, the average score for English users editing the Japanese edition (median 0.45, mean 0.44, sd 0.53) was marginally higher but not significantly different from the average score for Japanese users editing the Japanese edition (median 0.37, mean 0.43, sd 0.50, $p=0.72$).}

\section{Discussion}

The large differences in content observed between different editions of Wikipedia globally \cite{hecht2010} also applied to articles related to Okinawa despite the presence of Japanese and English speakers living on the island. Similarly, the small percentage of users editing multiple editions of Wikipedia \cite{hale-dphil-wikipedia} applied to this dataset as well. In many ways, Okinawa is a hard case: Japanese and English use very different writing systems, and Japanese users have consistently been observed to engage less with other-language content not only on Wikipedia \cite{hale-dphil-wikipedia}, but also on Twitter \cite{hale-chi2014}.

Nonetheless, this work has shed greater light on the selection of articles multilingual users in these two languages edit, the types of contributions they make, and the value of these edits. If further research confirms the patterns found in this \thing{} apply more broadly, then one key challenge for designers of multilingual platforms seeking to facilitate cross-language information exchange is increasing multilingual users' awareness of and contributions to related other-language content.
The multilingual users in this study were far less likely to edit articles in their non-primary languages that did not have corresponding articles in their primary languages despite these articles being more numerous. The large difference in content between languages applies not only to the sample used for this study, but also overall on Wikipedia \cite{hecht2010}, on Twitter \cite{hong2011}, and likely to most other user-generated content platforms. The challenge of making users aware of content available in their non-primary languages but not in their primarily languages is thus a challenge likely to be faced by designers of all multilingual user-generated content platforms.

Further research will be needed to understand the precise implications of design on user content selection, but it seems likely that the prominence of interlanguage links connecting related articles across languages on Wikipedia and the lack of other-language discovery tools are partially responsible for the narrow scope of multilingual users' edits in their non-primary languages. Currently, for example, there is no facility to search multiple language editions of Wikipedia simultaneously. So, if a user generally searches only in his primary language, that user may not discover the content that exists in another language if the content has no corresponding article in his primary language even if the user reads the other language. Very often the full text of articles in the Japanese edition includes an English translation of the article's concept, and likewise Japanese terms are often included in Japanese-themed articles in the English edition. If the search interface automatically checked a user's non-primary language editions when no matches were found in the user's primary language edition, that user might discover articles in his non-primary languages that have no article in his primary language.

Another possible design change would be to suggest articles related to a specific theme that exist in users' non-primary languages, but not in their primary languages. Such an approach could employ similar methods to those used here: gathering all articles linking to a given article and computing the PageRank scores or other methods like Latent Dirichlet Allocation (LDA) \cite{wikipediaLDA2008}. In practice, this might look very similar to the valuable SuggestBot \cite{cosley2007} tool, which can recommend articles for users to edit based on the articles the users have previously edited in one language. Currently, separate versions of SuggestBot run independently in multiple languages, and a potentially useful (although certainly non-trivial) step would be to extend SuggestBot to offer suggestions across user-selected (or inferred) languages for users who desire such suggestions. That is, based on the articles a user has previously edited in one language, the user could ask for articles in another language that either need work or that have no equivalent in the user's primary language.

A small, but dedicated group of users, who made large-sized edits in their primary languages, also edited articles in a second language. The edits in the users' non-primary languages were smaller in size, but were equally valued by the site's users persisting through subsequent revisions at a similar rate to the edits made by users editing only one language edition. An exploratory analysis of the edits users made in their primary and non-primary languages indicated that the differences in edit sizes were partially due to users making different types of edits in their primary and non-primary languages---multilingual users more frequently edited images and interlanguage links in their non-primary languages. In contrast, they made more maintenance and deletion/reversion edits in their primary languages.

Even if the edits in users' non-primary languages are smaller and of a different type, they still have value. The qualitative exploration of edits revealed many examples of users updating out-of-date information and correcting errors in their non-primary languages. Japanese users also frequently added or corrected relevant Japanese-language text in the English edition. There were also edits of addition and, occasionally,  translation into users' non-primary languages.

The Language Engineering team of the Wikimedia Foundation\footnote{\url{https://wikimediafoundation.org/wiki/Language_Engineering_team}}
has been actively developing an (open-source) content translation tool to help users translate content between different language editions of Wikipedia.\footnote{\url{https://www.mediawiki.org/wiki/Content_translation}}
While the integration of machine translation and bilingual dictionaries, the automatic conversion of article links, and the streamlined user-interface of the translation tool will no doubt assist would-be translators, this research indicates that helping users find articles they want to translate will be a major hurdle. Multilingual users in this study clearly made their largest-sized contributions in their primary languages, suggesting that platforms might be more successful in encouraging translation from users' non-primary languages to their primary languages rather than from users' primary languages to their non-primary languages. This would require surfacing content in users' non-primary languages that does not exist in the users' primary languages even while the users are viewing content in their primary languages.

Beyond translation, there are a range of contributions users can make on multilingual user-generated content platforms that require a varying level of cross-language proficiency. Offline data on multilingualism is imperfect and incomplete, but best estimates suggest around half of all humans speak two or more languages \cite{grosjean2010}. Thus, it might be possible to encourage far more users to contribute content in multiple languages on user-generated content platforms. Survey work suggests that Internet users consume content in multiple languages more frequently than they contribute content in multiple languages (not only on Wikipedia,\footnote{\wikisurvey2011} but also more generally online \cite{eurobarometer2011}).
The prevalence of image-related edits in this study and the apparent motivation images provide for cross-language linking in the blogosphere \cite{hale-msc} suggest multimedia content is a low-barrier entry point for increasing multilingual contributions from users. Designers of multilingual user-generated content platforms wishing to increase multilingual activity could specifically consider and optimize cross-language multimedia content related tasks and other low-barrier entry points for multilingual contributions as one possible way to increase the number of users contributing in multiple languages on their sites.

Adler et al.~\cite{adler2008-assigningtrust,adler2007,adler2008-measuring} advocate combining together the measures of edit size and edit persistence to form a reputation score for each user, which their WikiTrust work uses to predict the quality or trustworthiness of users' contributions to Wikipedia. If such a system evaluated the contributions of multilingual users separately for each language, most multilingual users would have low reputation scores in their non-primary languages due to the smaller sizes and smaller number of their edits while they would have large reputation scores in their primary languages. The analysis in this \thing{}, therefore, shows the importance of evaluating multilingual users holistically across their multiple languages to accurately measure their contributions to user-generated content sites.

Successfully combating the risk of fragmenting users and content too thinly across multiple languages on user-generated platforms requires a more advanced understanding of localization and internationalization than traditional principals such as translating interfaces to ``speak the user's language(s)'' \cite{nielsen1993}.
Multilingual users need to be specifically considered in site design. This involves not only making existing cross-language connections visible, but also designing for the discovery of related foreign-language content not available in users' primary languages.
The most active and dedicated users will reach across language boundaries online to contribute to other-language content, but the users in this study made their largest-sized contributions in their primary languages.
Thus, successful multilingual user-generated content platforms need to nurture both active, dedicated monolingual communities and also encourage multilingual users to discover other-language content and serve as bridges between monolingual communities. Further research should analyze the extent to which multilingual search, cross-language content recommendation, and optimized low-barrier entry points for multilingual contribution can help multilingual users better understand the differences in content between languages, encourage these users to transfer more information between different languages, and thereby enable wider access for all  users to the most interesting and important material that is not yet in their primary languages.

\section{Acknowledgments}
I would like to thank Eric T.~Meyer, Taha Yasseri, and Alolita Sharma 
as well as the anonymous  reviewers who provided helpful comments on previous versions of this article.

\bibliographystyle{acm-sigchi}

\end{document}